\documentclass{article}

\usepackage{arxiv}

\usepackage[utf8]{inputenc} 
\usepackage[T1]{fontenc}    
\usepackage{hyperref}       
\usepackage{url}            
\usepackage{booktabs}       
\usepackage{amsfonts}       
\usepackage{nicefrac}       
\usepackage{microtype}      
\usepackage{lipsum}
\usepackage[utf8]{inputenc}
\usepackage{amsmath,amssymb,amsfonts}
\usepackage{algorithmic}
\usepackage{textcomp}
\usepackage{underscore}
\usepackage{verbatim}
\usepackage{graphicx}
\usepackage{caption}
\usepackage{epstopdf}
\usepackage{subfigure}
\usepackage{makecell}
\usepackage{multirow}

\usepackage{algorithm}
\floatname{algorithm}{Algorithm}

\title{Adaptive In-network Collaborative Caching for Enhanced Ensemble Deep Learning at Edge}

\author{\uppercase{Yana Qin}$^{1,2}$, \uppercase{Danye Wu}$^{2}$, \uppercase{Zhiwei Xu}$^{1,2}$,\uppercase{Jie Tian}$^{3}$, and \uppercase{Yujun Zhang}$^{2}$\thanks{Corresponding author} \\
	$^{1}$College of Data Science and Application, Inner Mongolia University of Technology, Hohhot 100080, China\\
	$^{2}$Institute of Computing Technology, Chinese Academy of Sciences, Beijing 100190, China\\
	$^{3}$Facebook, New York 07102, USA
}

\begin{document}
\maketitle

\begin{abstract}
To enhance the quality and speed of data processing and protect the privacy and security of the data, edge computing has been extensively applied to support data-intensive intelligent processing services at edge. Among these data-intensive services, ensemble learning-based services can in natural leverage the distributed computation and storage resources at edge devices to achieve efficient data collection, processing, analysis. 

Collaborative caching has been applied in edge computing to support services close to the data source, in order to take the limited resources at edge devices to support high-performance ensemble learning solutions.
To achieve this goal, we propose an adaptive in-network collaborative caching scheme for ensemble learning at edge. First, an efficient data representation structure is proposed to record cached data among different nodes. In addition, we design a collaboration scheme to facilitate edge nodes to cache valuable data for local ensemble learning, by scheduling local caching according to a summarization of data representations from different edge nodes. 
Our extensive simulations demonstrate the high performance of the proposed collaborative caching scheme, which significantly reduces the learning latency and the transmission overhead.
\end{abstract}


\section{Introduction}
\label{sec:introduction}
With the breakthrough of Artificial Intelligence (AI), we are witnessing a booming increase in AI-based applications and services. The existing intelligent applications are computation intensive.  
To process a huge number of data in time at the edge of the network, edge computing has rapidly developed in recently. Edge computing~\cite{deng2020edge} takes a part of the resources and memory from the data center and puts it at the edge of the network to be closer to end users, which can reduce the network transmission delay, protect user's privacy and improve the network experience of end users. 

The rapid uptake of edge computing applications and services pose considerable challenges on networking resources. Fulfilling these challenges is difficult due to the conventional networking infrastructure. 
The extensive application of deep neural network models at edge makes this problem more serious.  
Neural network models learn relationships among a huge number of training data. Meanwhile, this type of complex nonlinear models is sensitive to initial conditions, both in terms of the initial random weights and in terms of the statistical noise in the training data. This stochastic nature of the learning algorithm means that a neural network model is trained, it may learn a new group of features from inputs to outputs, which have different performance in practice. 

It is the ensemble learning~\cite{zhou2019edge,chen2019artificial} that provide a feasible way to handle the variance of a single neural network model. Ensemble learning schemes train multiple models and combine their outputs to alleviate the variance of a single model. We can achieve a high-quality neural network model by combining different neural network models to an ensemble result. 

Although ensemble learning can enhance the capability of edge deep learning, there is no performance improvement if the similar individual sub-models are combined~\cite{zhou2009ensemble}. Tumer et al.~\cite{tumer1995theoretical} analyzed simple soft vote ensemble methods by decision boundary analysis, revealing the importance of difference among sub-models. The same conclusion is also applicable to other ensemble methods. However, it is not easy to generate individual sub-models with high diversity. A significant challenge comes up that sub-models are obtained on similar training data, so they are often highly correlated. 

To make all sub-models different from each other, we target on an adaptive collaborative caching scheme to guarantee sub-models learning different data and consequently being diverse. With comprehensive study of collaborative caching at edge, we propose a compact recording structure of the cached data, maximizing the difference among the data cached in different nodes. Besides the caching locality considered in the conventional caching schemes~\cite{yuan2018toward}, it is also important to consider the diversity of the data cached to train sub-models. In this way, we improve the performance of the ensemble learning. Our main contributions are summarized as follows: 

\begin{enumerate}
	\item To efficiently record cached data among collaborative nodes, we study data recording and exchanging among edge nodes, and introduce a compact representation structure, Combinable Counting Bloom Filter (CCBF). 
	\item We design an adaptive collaborative caching scheme based on CCBF, based on which a high-performance ensemble deep learning at edge is proposed. 
	\item Comprehensive evaluation of the proposed scheme is performed on NS-3 based simulations over real-world deep learning models and data. 
\end{enumerate}

In Section~\ref{sec:related}, we study the related work. A compact data structure to collect the information of the cached data is present in Section~\ref{sec:CCBF}. Base on this data structure, we propose an adaptive collaborative caching scheme for ensemble learning in Section~\ref{sec:cache}. In detail, we exchange the information of the cached data, elevate these data to learn local knowledge, and achieve a high-performance ensemble result. The performance of our design is evaluated in Section~\ref{sec:experiment}. Finally, we conclude the work in Section~\ref{sec:con}.

\section{Related Work}
\label{sec:related}
In recent years, ensemble learning at the edge has been used in all kinds of applications~\cite{wang2018edge,yang2019federated}. Meanwhile, due to the storage limitation of each edge node, edge nodes always collaborate in data collection and model training~\cite{recht2011hogwild,zhang2015deep,wen2017terngrad}. 

\subsection{Ensemble Learning}
We investigate relevant works in recent years, and we find the performance can be effectively improved with the introduction of ensemble mechanism. To optimize the determination process of deep classification model structure and the combination of multi-modal feature abstractions, Yin et al.~\cite{yin2017recognition} proposed multiple-fusion-layer based ensemble classifier of stacked auto-encoder (MESAE) for recognizing emotions, in which deep learning is used for guiding autoencoder ensemble. Moreover, based on the assumption that different convolutional neural network (CNN) architectures learn different levels of semantic representations, Kumar et al.~\cite{kumar2016ensemble} developed a new feature extractor by ensembling CNNs that were initialized on a large dataset of natural images. Experiment showed that the ensemble of CNNs can extract features with a higher quality, compared with traditional CNNs. Xiao~\cite{xiao2019svm} proposed an ensemble learning method to improve the robustness in traffic incident detection. Galicia et al.~\cite{galicia2019multi} presented ensemble models for forecasting big data time series. Liu et al.~\cite{liu2017ensemble1} applied ensemble convolutional neural network models with different architectures for visual traffic surveillance systems. Liu et al.~\cite{liu2017ensemble} designed an ensemble transfer learning framework which used AdaBoost to adjust the weights of the source data and target data, this method achieved good performance on UCI datasets when the training data are insufficient. Chen et al.~\cite{chen2018ensemble} proposed an ensemble network architecture for deep reinforcement learning, in order to solve the problem that existing ensemble algorithms in reinforcement learning. 

\subsection{Collaborative Caching}
Collaborative caching has been applied in the ensemble learning field to collect sufficient data for sub-models training and high-quality ensemble result achievement. Amer et al.~\cite{amer2020caching} stated the role of wireless caching in low-latency wireless networks and characterized the network average delay on a per request basis from the global network perspective. Li et al.~\cite{li2019collaborative} proposed a cache-aware task scheduling method in edge computing. First, an integrated utility function is derived with respect to the data chunk transmission cost, caching value and cache replacement penalty. Data chunks are cached at optimal edge servers to maximize the integrated utility value. After placing the caches, a cache locality-based task scheduling method is presented. Chien et al.~\cite{chien2020q} proposed a collaborative cache mechanism in multiple Remote Radio Heads (RRHs) to multiple Baseband Units (BBUs). In addition, they use Q-learning to design the cache mechanism and propose an action selection strategy for the cache problem. Through reinforcement learning to find the appropriate cache state. Ndikumana et al.~\cite{ndikumana2017collaborative} proposed collaborative cache allocation and computation offloading, where the MEC servers collaborate for executing computation tasks and data caching. Tang et al.~\cite{tang2019using} proposed caching mechanisms in collaborative edge-cloud computing architecture, which can implement the caching paradigm in cloud for frequent n-hop neighbor activity regions. Khan et al.~\cite{khan2020lochip} proposed reversing the way in which node connectivity is used for the placement of content in caching networks, and introduce a Low-Centrality High-Popularity (LoCHiP) caching algorithm that populates poorly connected nodes with popular content. Wei et al.~\cite{wei2019automating} presented a system that automatically parallelizes serial imperative ML programs on distributed caching. The system makes a static dependence analysis to determine when dependence-preserving parallelization is effective, and maps a computational process to a distributed schedule.

\subsection{Problems and Our Insight}
Although existing caching approaches can facilitate ensemble learning, enhanced collaborative caching should be studied. Actually, the ensembling learning process is highly related to different sub-models.  As analyzed in Tumer et al.~\cite{tumer1995theoretical}, if the sub-models are independent of each other,  the error of ensemble learning will be reduced. If each sub-model is correlated with all others, the error of ensemble learning becomes larger. This analysis clearly reveals the importance of different sub-models~\cite{zhou2009ensemble}.

To differentiate all sub-models as much as possible, we take different data to train various sub-models and improve the performance of ensemble learning: (1) An efficient way to record the cached data items is highly required. (2) Exchange the record of the cached data among different edge nodes. (3) Schedule the edge caching according to the records and train different sub-models for high-quality ensemble learning.

\section{Composable Counting Bloom Filter}
\label{sec:CCBF}
To make the valuable data remain on edge nodes, we need to exchange the compact records of the cached data. However the most popular compact recording method, Counting bloom filter (CBF), only support inserting, deleting and query on data, and cannot support the combination operation of multiple filters which will be used to summarize the exchanged compact records of the cached data. Therefore, to meet the aforementioned requirement, we introduce a combinable Counting Bloom Filter (CCBF) in this section.  

\subsection{Design Structure}
To support the dynamic update of the record of the cached data, as well as the combination of multiple compact records of the cached data, we design a new structure for the proposed CCBF on the basis of basic bloom filter.
We can combine multiple basic bloom filters by performing bitwise OR on these filters, but cannot combine CBFs in the similar way since CBF aggregates the information of the inserted data into its counters.
Based on the above observation, we can stack several basic bloom filters to build a counting bloom filter which can support updating and merging operations simultaneously.

\begin{figure}[tb]
	\centering
	\includegraphics[width=0.5\textwidth]{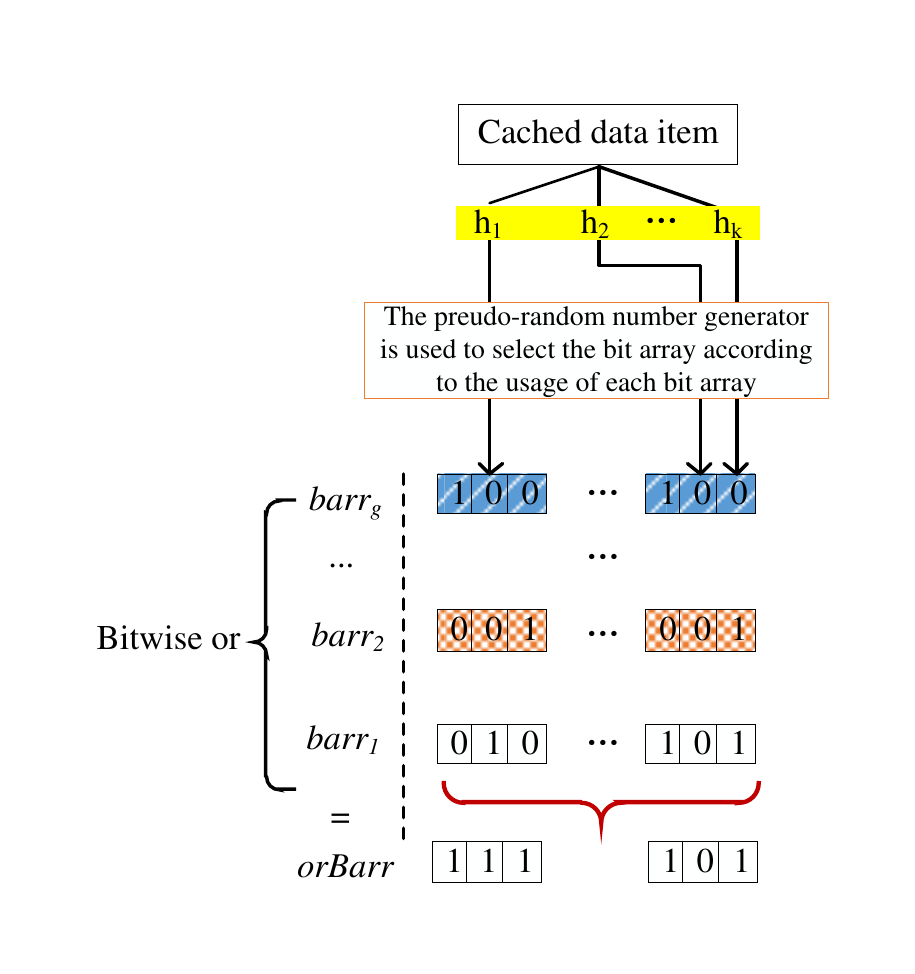}
	\caption{Structure of Composable Counting Bloom Filter}
	\label{Design Structure}
\end{figure}

The detailed structure of CCBF is shown in Figure~\ref{Design Structure}. The CCBF has $k$ hash functions ($h_{1}, h_{2}, \cdots, h_{k}$) and consists of the following two components:

\begin{enumerate}
	\item  \textit{G bit arrays ({$barr_{i}$}, $i=1,2,...,g$)}: The bit arrays used to replace the counter arrays in CBF to support counting operations of the inserted data items, the size of which is equal to $m$. $g$ is set based on the requirement of counting.
	\item \textit{orBarry}: The aggregation result of g bit arrays by performing bitwise OR, which is used to enhance the the query efficiency and facilitate the data caching among edge nodes.
\end{enumerate}

CCBF not only supports the insert, query and delete of items, but also supports the combination of multiple CCBFs. These operations will be described one by one in the next section.

\subsection{Related Operations}
According to the needs of exchanging and updating cached data records, we have implemented the functions of inserting, querying, deleting and combination in CCBF. In this part, we introduce the related operations.

\subsubsection{Inserting}
To insert a data item into CCBF, we use the pseudo-random integer generator to generate a random matrix to perform an efficient and non-repeated insert operation. The specific inserting operation is shown in Algorithm~\ref{alg:CCBF.insert} and the procedures are shown as follows:

\begin{enumerate}
	\item A random matrix($matrix[g][m]$) of size $g\times m$ is constructed by using a pseudo-random integer generator with different seeds on different columns. For each column, the value of each cell is different from the others, and belongs to a range from $1$ to $g$.
	\item Hash the cached data $k$ times to get $k$ hash results($\{p_{j}\}$) (line 3).
	\item Use the RandChoice function to  searches the $p_{j}-{th}$ column of $matrix[g][m]$ for the next available bit array, according to the number of arrays whose $p_{j}-{th}$ cells are used (line 4).
	\item Set the $p_{j}-th$ cell in $barr_{i}$ to be $1$ and update $orBarr$ (line 5-7).
\end{enumerate}

Notice that we will check If the bit arrays used to record the $k$ hash results meets the following condition:
\begin{eqnarray} 
	\left\{\begin{matrix}
		\forall barr_i[p_{j}]=1\\s.t.j\in{1,2,...,k}
	\end{matrix}\right.
\end{eqnarray}
It means that this data item has been inserted, and this insert operation will be abandoned. 

\begin{algorithm}[h]
	\caption{CCBF.insert($d$)}
	\begin{algorithmic}[1]
		\REQUIRE $d$
		\STATE $barrSet=CCBF.GetAllBarr();$
		\FOR {$j=0$ to $CCBF.k-1$}	
		\STATE  $p_{j}=Hash_{j}(d);$
		\STATE  $barr_{i}=RandChoice(barrSet, p_{j}, Counter(p_{j})); $
		\\//Function RandChoice searches the $p_{j}-{th}$ column of $matrix[g][m]$ for an available bit array, according to the number of arrays whose $p_{j}-{th}$ cells are used;
		\STATE $barr_{i}[p_{j}]=1;$
		\ENDFOR  
		\STATE $Update$ $orBarr();$ 
		\end {algorithmic}
		\label{alg:CCBF.insert}
		\end {algorithm}
		
		\subsubsection{Querying}
		We include an additional array $orBarr$ to support efficient membership query operations, which can directly check whether the corresponding cells of orBarr are set to be $1$. The specific query operation is shown in Algorithm~\ref{alg:CCBF.query}:
		
		\begin{enumerate}
			\item Hash the queried data $k$ times to get $j$ hash results($p_{j}$) (line 2).
			\item Check if $orBarr[p_{j}]$ is 1. If it is 1, the data is inserted before. Conversely, there is no corresponding data (line 3-7).
		\end{enumerate}
		
		\begin{algorithm}
			\caption{CCBF.query($d$)}
			\begin{algorithmic}[1]
				\REQUIRE $d$;
				\ENSURE Query result (true or false);
				\FOR {$j=0$ to $CCBF.k-1$}
				\STATE $p_{j}=Hash_j(d)$
				\IF {$CCBF.orBarr[p_{j}] \neq 1$}
				\RETURN $false$
				\ENDIF 
				\ENDFOR 
				\RETURN $true$
				\STATE End
			\end{algorithmic}
			\label{alg:CCBF.query}
		\end{algorithm}
		
		\subsubsection{Deleting}
		The calculation process of the delete operation is similar to that of the insert operation. The key operation steps are briefly listed as the follows:
		\begin{enumerate}
			\item Confirm whether the item exists in this CCBF by performing a query operation on this item (see ~\ref{alg:CCBF.query}).
			\item Locate the bit arrays used in the last inserting operation according to the random matrix $matrix[g][m]$.
			\item Clear the corresponding cells in these bit arrays, and update $orBarr$.
		\end{enumerate}
		
		\subsubsection{Combination}
		CCBF does not only support inserting, query and deleting data items, but also supports the combination of multiple CCBFs. The combination operation of multiple CCBFs is equivalent to combine the data items inserted in these CCBFs. 
		The random matrix $matrix[g][m]$ generated in CCBF can ensure that the bit array selected in a fixed sequence, and thus the repeated data inserting can be neglected. The specific combination operation is shown in Algorithm~\ref{alg:CCBF.combination} and the combination procedure is listed as the follows:
		
		\begin{enumerate}
			\item Determine whether the number of items in the compacted representation after merging has exceeded the capacity of CCBF ($n$) (line 1-3).
			\item Combine bit arrays one by one by bitwise OR (line 6-13).
		\end{enumerate}
		
		\begin{algorithm}
			\caption{CCBF.combine($other$)}
			\label{alg:CCBF.combination}
			\begin{algorithmic}[1]
				\REQUIRE Two CCBFs, $CCBF$ and $another$;
				\IF {$CCBF.Size()+other.Size() >  CCBF.n$}
				\RETURN $error$;
				\ENDIF 
				\STATE $barrSet=CCBF.GetAllBarr()$;
				\STATE $otherbarrSet=other.GetAllBarr()$;
				\FOR {$j=0$ to $CCBF.g-1$}
				\STATE $barr=barrSet.first$;
				\STATE $otherbarr=otherbarrSet.first$;
				\STATE $barr.bitwiseOr(otherbarr)$	;
				\STATE $barrSet.Remove(barr)$;
				\STATE $otherbarrSet.Remove(otherbarr)$;   
				\ENDFOR 
				\STATE $CCBF.orBarr.bitwiseOr(other.orBarr)$;
			\end{algorithmic}
		\end{algorithm}
		
CCBF can record the cached data in each edge node through a compact way. By inserting the data items in CCBF and exchanging and combining CCBFs among edge nodes, the cache of each edge node can be scheduled to store diverse data, which facilitates obtain different sub-models on these data and achieve an high-performance ensemble learning result.

\section{Ensemble Learning Based on Collaborative Caching}
\label{sec:cache}
In this section, we propose an edge ensemble learning scheme based on adaptive collaborative caching. In this scheme, CCBF is used to record the cached data information, so as to realize the exchange and collection of cache information between edge nodes. Also, the scheme can reasonably schedule the cache according to the data distribution, and support the sub-models learning and final ensemble learning of each edge node.

\begin{figure*}
	\centering
	\includegraphics[width=0.9\textwidth]{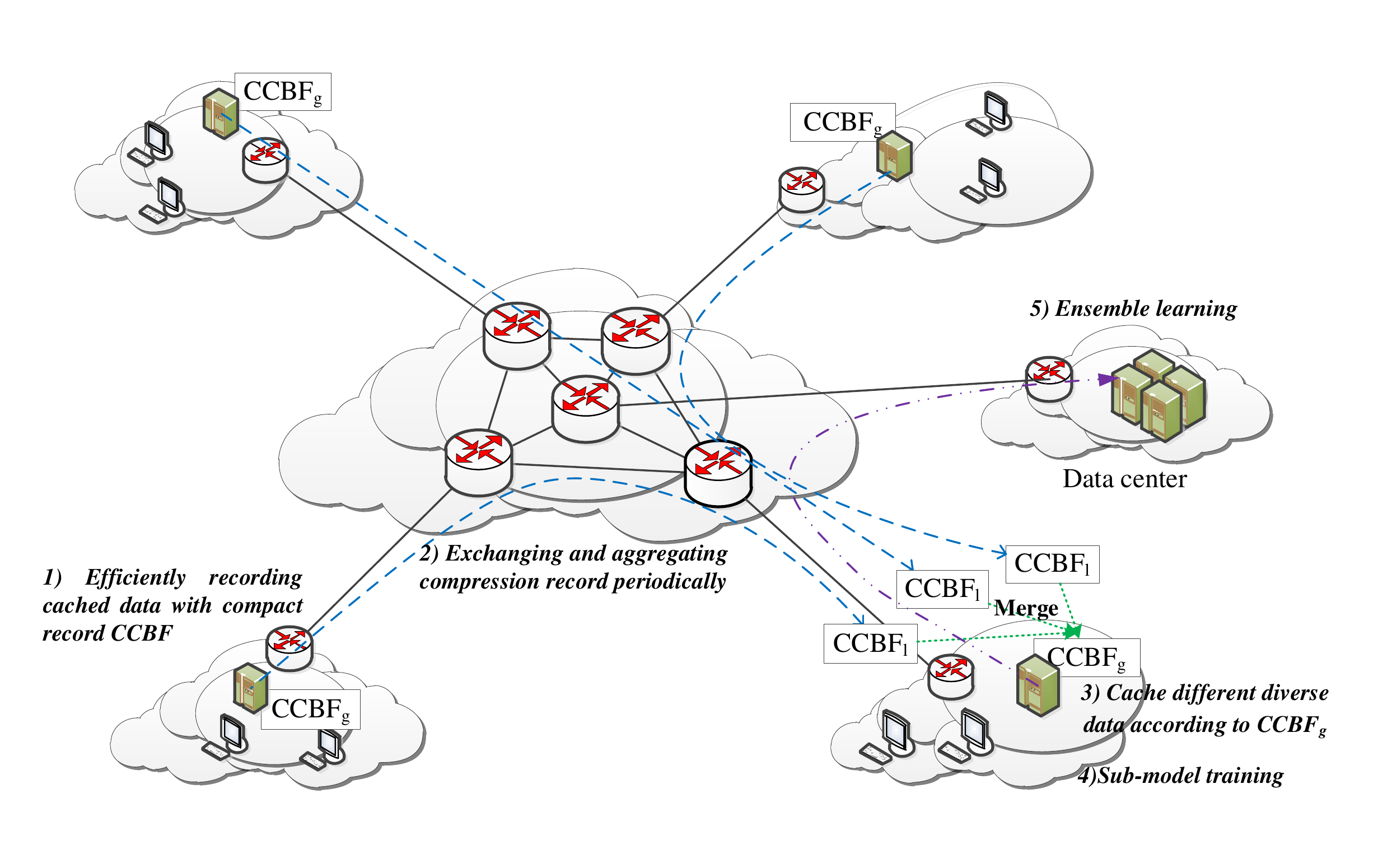}
	\caption{Adaptive In-network Collaborative Caching Process}
	\label{scheme}
\end{figure*}

\subsection{Learning Strategy}
\label{sec:learningstrategy}
In order to improve the performance of ensemble learning at edge, we first study the ensemble learning process. The decision boundary analysis results of the simple soft voting ensemble method are as follows:

To remain simple, it is assumed that all sub-models have the same error rate. We use $\theta$ to describe the relationship between different sub-models. The expectation error of ensemble learning is
\begin{eqnarray} 
	\label{equal1}
	\bar{err}\left ( H \right )=\frac{1+\theta \left ( n-1 \right )}{n}err_{i}(h_{i}),i=1,2,...,n
\end{eqnarray}
where$\ err_{i}(h_{i})$ is the expectation error rate of a sub-model, and $n$ is the size of the ensemble scale. Formula~\eqref{equal1} shows that if the sub-models are independent of each other, \emph{i.e.}, $\theta=0$, the error of ensemble learning will be reduced by $n$ times. If each sub-model is correlated with all the others, \emph{i.e.}, $\theta=1$, the performance of ensemble sub-models will not be effectively improved. This analysis clearly reveals the importance of different sub-models in ensemble learning, and the same conclusion applies to other ensemble approaches~\cite{zhou2009ensemble}. In edge ensemble learning scenarios, different edge nodes often deploy similar models, build sub-models by learning the data around the edge nodes, and eventually form an ensemble model by distributing the sub-models on different nodes by the central node. For this case, it is necessary to provide different data for different edge nodes to get different training results of sub-models, so as to achieve a more accurate ensemble model.

\subsection{Leveraging Collaborative Caching to Facilitate Ensemble Learning Process}
According to the analysis in Section~\ref{sec:learningstrategy}, we leverage a collaborative caching scheme to support different sub-models learning of ensemble learning. In detail, the process can be divided into the following five phases.

\subsubsection{Efficiently Recording the Cached Data on a Edge Computing Node}
Data is collected from neighboring end devices, which will be cached and recorded in the compact way by CCBF (see Section~\ref{sec:CCBF}).

\subsubsection{Exchanging Compact Representation of Cached Data with Neighbors}
The compact representation (CCBF) of the cached data is exchanged among neighbors in a range, which adapts to the performance improvement of the sub-model training in step 4. Once a neighbor receives a representation from an interface, this representation will be stored, with a name of $CCBF_l$, where $l$ is the id of the corresponding interface. In addition, the representation is combined with an aggregated representation on this neighbor, $CCBF_g$, and gain a global view about the data cached in the neighbors, which will be used to guide the neighbor to cache various data received subsequently.\textit{ Notice that, although more data from a big range of neighbors can benefit the training performance of the sub-models, the collaboration in a big range will cause larger communication overhead, and thus our design make the collaborative range adapt to practical sub-model training results. }

\subsubsection{Caching Different Data among Neighbors}
When the neighbor requests to cache some data, it first need to check whether this data has already existed in the neighbors. It queries the $CCBF_g$ that represents the global view about the data cached in the neighbors. If the record of this data is found in $CCBF_g$, which indicates that the data has already existed in the cache of other neighbors, the data no longer need be stored in the neighbor's cache. And if this record does not exist in the $CCBF_g$, indicating that the caches of other neighbors do not contain this data, this data can be added to the cache and a record is added to the corresponding compaction record. The above operation ensures that different data can be cached at neighbors for training different sub-models in ensemble learning, while reducing the communication overhead by collaborative caching.

\subsubsection{Sub-model Training}
The data cached on one node is used to train the local sub-model. When the local data is not enough to make the sub-model converge, we need to enlarge the collaborative range by requesting differentiated data from the other edge nodes.
We compare the cached data records from different neighbors obtained in step 2, $CCBF_l$, with the local cache record by performing merge on the $orBarr$ of different $CCBF_l$ from different neighbors, obtain the required data compact representation $\hat{orBarr}$ and send it to the corresponding edge node. When the corresponding edge node receives the request, it queries the local cache according to $\hat{orBarr}$ and returns the differentiated data to the requesting node. After the requesting node receives the data, it caches the data and updates $CCBF_l$ and $CCBF_g$, then inputs the data into the sub-model for training. The procedures are repeated until the sub-model converges.

\subsubsection{Enhanced Ensemble Process}
The ensemble method obtains the result by attaching different weights to the output result of each sub-model. The ensemble output result $H\left (x \right)$ is:
\begin{eqnarray}
	H\left ( x \right )=\sum_{i=1}^{n}\omega _{i}h_{i}\left ( x \right )
\end{eqnarray}
where $\omega _{i}$ denotes the weight of $h_{i}$, usually with constraints $\omega_{i}\geq 0$ and $\sum_{i=1}^{n}\omega_{i}=1$.
These parameter weights from the sub-model are uploaded to the central node, which conducts ensemble learning in an enhanced way. Specifically, for $n$ sub-models ($h_{1},...,h_{n}$), the following method is adopted for ensemble learning. 

Suppose the output of each sub-model can be written as the true value plus an error term:
\begin{eqnarray}
	h_{i}\left ( x \right )=f\left ( x \right )+\epsilon_{i} \left ( x \right ),i=1,...,n
\end{eqnarray}

The ensemble error can be expressed as~\cite{perrone1993networks}:
\begin{eqnarray}
	\begin{aligned}
		\hat{err}\left ( H \right ) &= \int \left ( \sum_{i=1}^{n}\omega _{i}h_{i}\left ( x \right )-f\left ( x \right ) \right )^{2}p\left ( x \right )dx \\&= \int \left ( \sum_{i=1}^{n}\omega _{i}h_{i}\left ( x \right )-f\left ( x \right ) \right ) \times \\& \left ( \sum_{j=1}^{n}\omega _{j}h_{j}\left ( x \right )-f\left ( x \right ) \right ) p\left ( x \right )dx \\&= \sum_{i=1}^{n}\sum_{j=1}^{n}\omega _{i}\omega _{j}C_{ij}
	\end{aligned}
\end{eqnarray}

where $p\left (x \right)$ is the input distribution, $\epsilon_{i}$ is an error term, and 
\begin{eqnarray}
	C_{ij}=\int \left ( h_{i}\left ( x \right )-f\left ( x \right ) \right )\left ( h_{j}\left ( x \right )-f\left ( x \right ) \right ) p\left ( x \right )dx
\end{eqnarray}

The optimal weight can be solved in the following ways:
\begin{eqnarray}
	\omega =\underset{\omega }{argmin}\sum_{i=1}^{n}\sum_{j=1}^{n}\omega _{i}\omega _{j}C_{ij}
\end{eqnarray}

By means of the Lagrangian multiplier, we get $\omega_{i}$ is
\begin{eqnarray}
	\omega_{i}=\frac{\sum_{j=1}^{n}C_{ij}^{-1}}{\sum_{k=1}^{n}\sum_{j=1}^{n}\omega _{i}\omega _{j}C_{kj}^{-1}}
\end{eqnarray}

\section{PERFORMANCE EVALUATION}
\label{sec:experiment}
In this section, we conduct experimental simulations of two learning models on four different datasets to evaluate the performance of the proposed collaborative caching scheme for ensemble learning at edge.

\subsection{Implementation}
We evaluate the performance of our adaptive collaborative caching scheme on Ns-3 platform~\cite{carneiro2010ns}. It's a modular, programmable, extensible, open, open-source, community-supported simulation framework for computer networks. We connect neural networks library OpenNN (Open Neural Networks Library)~\cite{lopez2014open} to Ns-3 for experimental simulations. OpenNN is an open-source neural network library to facilitate the building of neural networks. It has found a wide range of applications, which include function regression, pattern recognition, time series prediction, optimal control, optimal shape design or inverse problems. All simulations are performed on a local machine, equipped with an Intel Core i7, 3.4G CPU and 16G RAM, running Ubuntu 16.04 with kernel version 3.19. 

\begin{figure*}[t]
	\centering
	\includegraphics[width=0.98\textwidth]{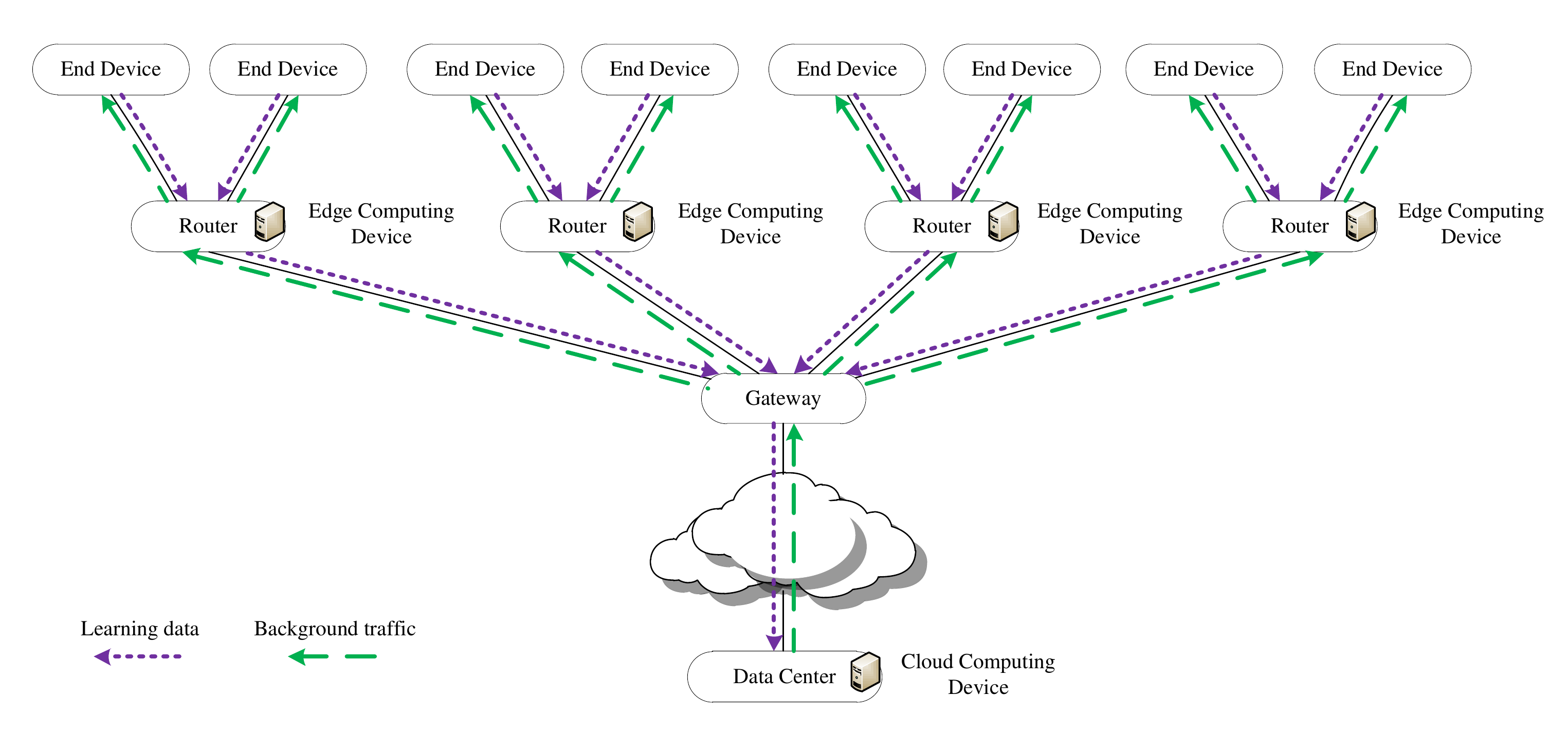}
	\caption{Simulation topology}
	\label{topology}
\end{figure*}

As shown in Figure~\ref{topology}, we use a general topology for edge networks. It includes a remote data center, a gateway node, 4 edge computing nodes and 8 end devices, connected by Gigabit links. The cache size of edge computing nodes is 2,000. Edge computing nodes can cache data and efficiently record cached data and perform cached data computational tasks.

End devices generate the learning data of models, and send data to edge computing nodes. After receiving the data,  edge computing nodes first carry out data caching and efficient recording, then use different data of cooperative caching to train sub-models, and finally send the training results of sub-models to the data center for ensemble learning.

The data center generates background traffic data and sends data to edge computing nodes.  After receiving the data, edge computing nodes carry out data caching and send background traffic data to end devices.

Two types of learning models are deployed on these edge nodes and the data center server. Correspondingly, four different datasets (D1, D2, D3 and D4) are used to compare the proposed adaptive collaborative caching scheme (C-cache)  with two baseline schemes implemented as the follows: 
\begin{itemize}
	\item  \textbf{Centralized:} In the model training process, the data center requests training data from edge nodes to train ensemble learning models on the server. 
	\item  \textbf{P-cache: }A caching mechanism in collaborative edge-cloud computing architecture~\cite{tang2019using} is proposed, in which edge computing nodes periodically request cached data for sub-models learning, and data center server performs ensemble learning.
\end{itemize}

To evaluate the performance of the collaborative caching scheme, we use four datasets to learn. Specifically, two text datasets (D1, D2) to train MLP model. To train VGG model, we apply an image dataset (D3) of tiger faces and a dataset (D4) of human faces. 

\begin{itemize}
	\item Covertype dataset~\cite{gama2003accurate} (D1): This dataset includes forest vegetation types in the Roosevelt National Forest. There are four types of soil, corresponding to seven types of vegetation. 581,012 data items forest vegetation are shown in four soil types. The numbers of data items for different soil types are not even. The number in type 4 is fewer than 3,000, and is almost 10,000 for type 5. The number of any other types is larger than 10,000.  
	\item Healthy Old People dataset~\cite{wickramasinghe2016sequence} (D2):  Sequential motion data from 14 healthy older people aged 66 to 86 years old using sensors for the recognition of activities in clinical environments. Participants were allocated in two clinical room settings (S1 and S2). S1 (Room1) and S2 (Room2) are set with different number and location of sensor receiver. The number of data items is 75,128, which are evenly categorized into 6 different behaviors. 
	\item Atrw Reid-tigerface dataset~\cite{schneider2020similarity} (D3): The images of Atrw Reid-tiger face are captured. After clipping, the image resolution is adjusted to 128 × 128. Each one of 500 tigers has 10 photos. According to the areas of activities, the Russian Far East region and the northern region of India, the dataset is separated into two scenarios. 
	\item Casia-face dataset~\cite{sarhan2017multimodal} (D4): The face images of human face are captured. After clipping, the image resolution is adjusted to 128 × 128. Each one of 500 persons has 10 face pictures. According to the Angle of the photograph taken, the position of the front and the rhombic 45 degrees, the dataset is separated into two scenarios. 
\end{itemize}

We implement the following two learning models, among which Adam algorithm~\cite{zhang2020online} is used, which can adaptively adjust the learning rate of the model: 

\begin{itemize}
	\item Multilayer Perceptron (MLP) model. MLP is a feedforward artificial neural network that maps multiple input data to outputs. The layers of MLP are fully connected. We implement a six-layer MLP model, including an input layer, four hidden layers, and an output layer. 
	\item Visual Geometry Group Network (VGG) model. VGG is a deep convolutional neural network used in computer vision. Our implementation includes 5 convolutional blocks with each consisting of 2-4 convolutional layers. For the five convolutional blocks, each of their layers contains 64-128-256-512-512 convolution kernels respectively. 
\end{itemize}

\subsection{Evaluation Metrics}
To evaluate the performance of adaptive collaborative caching scheme for ensemble learning at edge, we use three metrics: hit ratio, latency and accuracy.  

\subsubsection{Hit Ratio of Collaborative Caching} 
The concept of hit ratio is defined for any two adjacent level of memory in memort hierarchy. The performance of cache is measured in terms of hit ratio. If a data item requested by edge computing nodes is found in the cache, it is called a hit. Hit ratio is the number of hit data divided by total data items and consists of local hit ratio, global learning hit ratio and global background hit ratio.
\begin{enumerate}
	\item Local learning hit ratio ($LLR_{hit}$) represents how many local cached data can be used to train a sub-model. For example, if 30 pieces of data items in the local cache can be used for training the model, and the total amount of cached data items is 100, then the $LLR_{hit}$ is 30/100=0.3. $LLR_{hit}$ is the ratio of the learning data in the local cache and the overall cached data:
	\begin{eqnarray}
		LLR_{hit}=\frac{N_l}{N_c}
	\end{eqnarray}
	
	where $N_l$ is the number of data items for training a sub-model in the local cache, and $N_c$ is the total amount of locally cached data items.
	\item Global learning hit ratio ($GLR_{hit}$) represents how many data items in edge nodes can be used to train sub-models. $GLR_{hit}$ is the ratio of the learning data items in the global cache and the overall cached data items, calculated as:
	\begin{eqnarray}
		GLR_{hit}=\frac{N_{g}}{N_{gc}}
	\end{eqnarray}
	
	where $N_g$ is the number of data items for training sub-models in the global cache and $N_{gc}$ is the total amount of globally cached data items.
	
	\item Background hit ratio ($R_{hit}$) is the ratio of the background traffic data items in the global cache and the overall cached data items. Background traffic refers to the flow of data packet exchange between application program and network periodically or intermittently when there is no specific interaction. $R_{hit}$ calculated as:
	\begin{eqnarray}
		R_{hit}=\frac{N_b}{N_{gc}}
	\end{eqnarray}
	
	where $N_b$ is the number of background traffic data items.
\end{enumerate}

\subsubsection{Transmission overhead and Learning Latency} 
\begin{enumerate}
	\item The data transmission overhead is the size of data requested to support model training among edge computing nodes.
	\item Learning latency is a time period how long the training model converges.
\end{enumerate}

\subsubsection{Learning Accuracy}
The accuracy ($Acc$) of model training on a dataset is defined as: 
\begin{eqnarray}
	Acc=\frac{TP+TN}{TP+FN+FP+TN}
\end{eqnarray}
where True Positive (TP) is the number of positive items that are correctly classified, False Positive (FP) is the number of positive items that are incorrectly classified, False Negative (FN) is the number of negative items that are incorrectly classified, and True Negative (TN) is the number of negative items that are correctly classified.

\begin{figure}
	\centering
	\subfigure[$LLR_{hit}$ during training MLP on D1]{\includegraphics[width=0.36\textwidth]{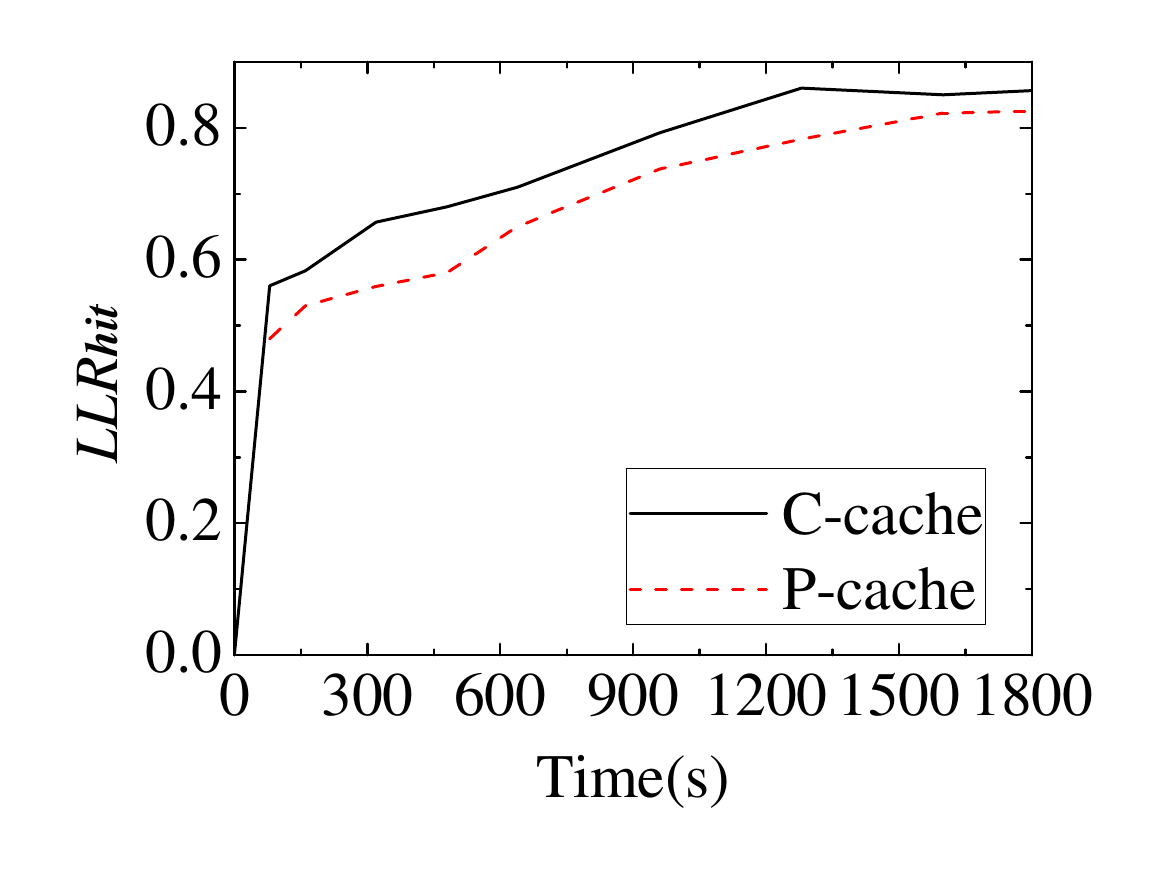}}
	\subfigure[$LLR_{hit}$ during training MLP on D2]{\includegraphics[width=0.36\textwidth]{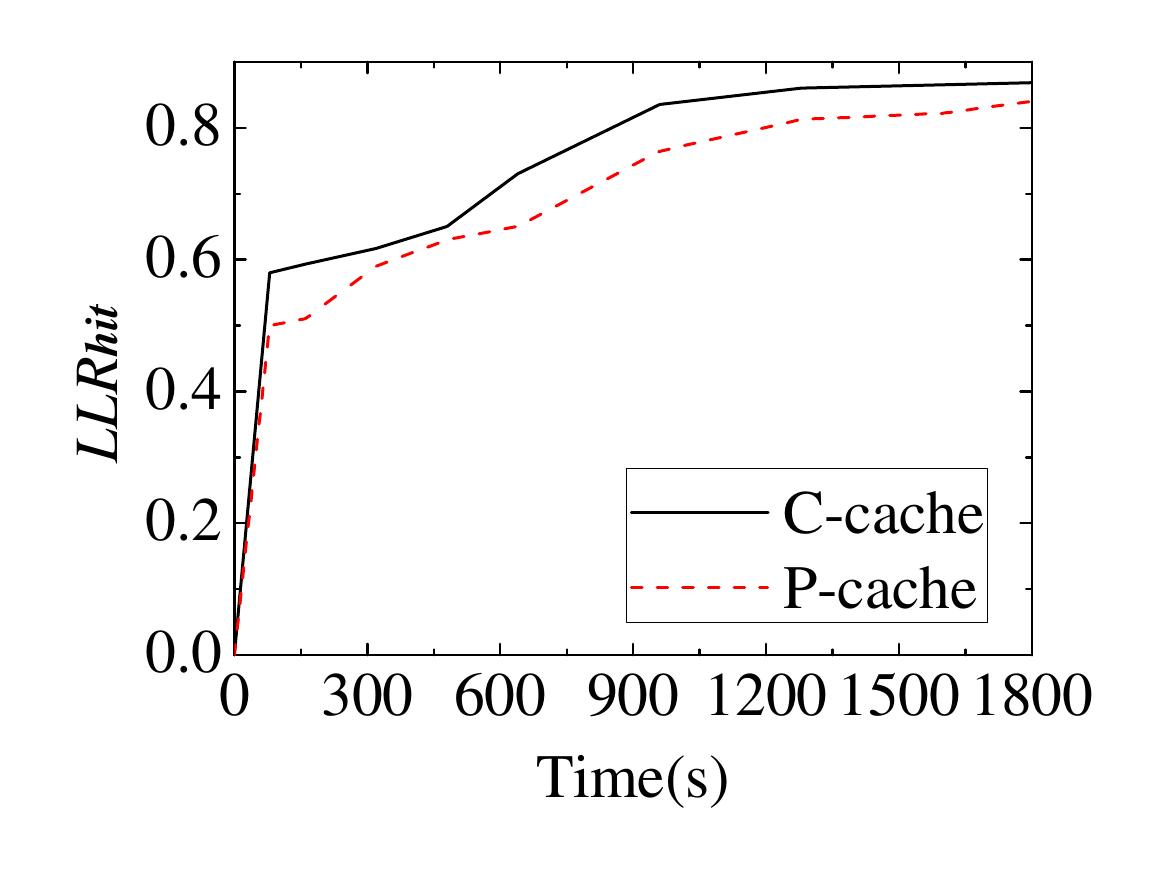}}
	\caption{Local hit ratio of MLP}
	\label{Local hit ratio1}
\end{figure}

\subsection{Evaluation Results}
\subsubsection{Hit Ratio of Collaborative Caching}
\begin{figure}
	\centering
	\subfigure[$LLR_{hit}$ during training VGG on D3]{\includegraphics[width=0.36\textwidth]{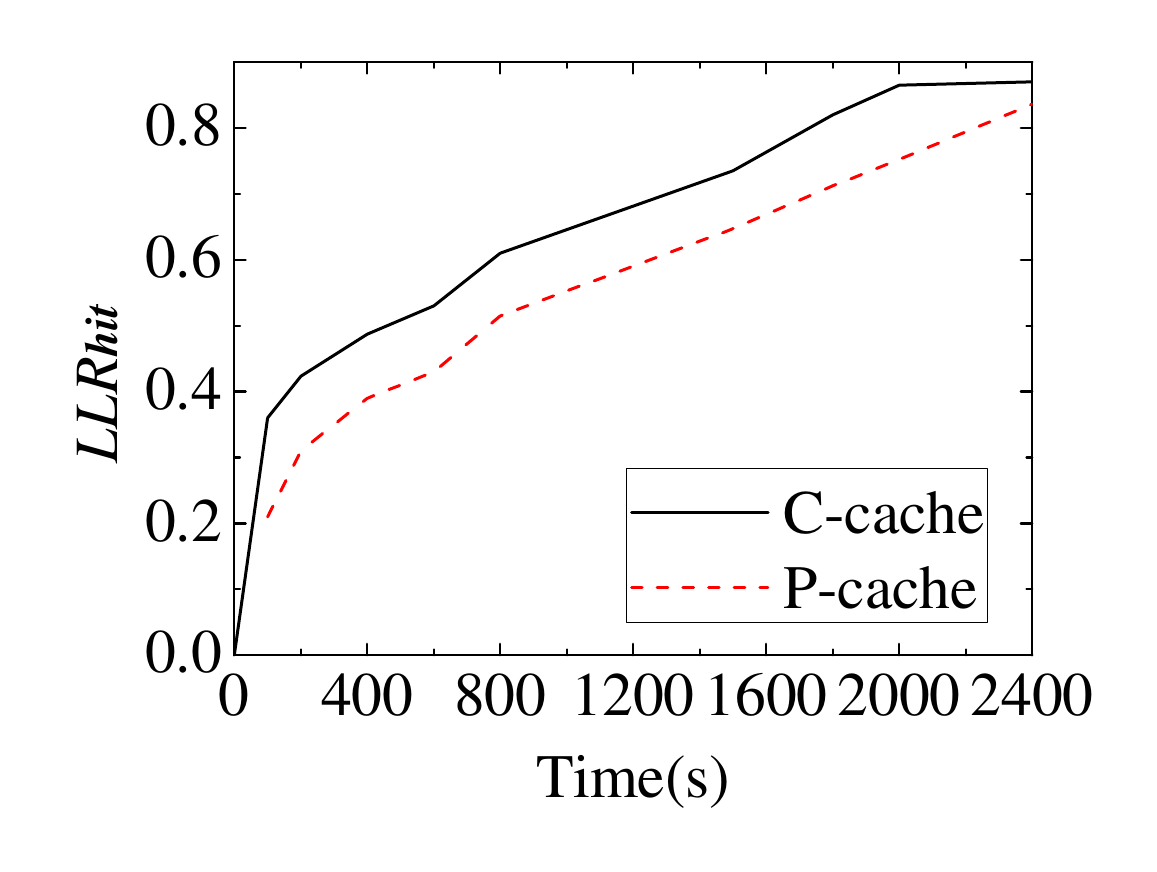}}
	\subfigure[$LLR_{hit}$ during training VGG on D4]{\includegraphics[width=0.36\textwidth]{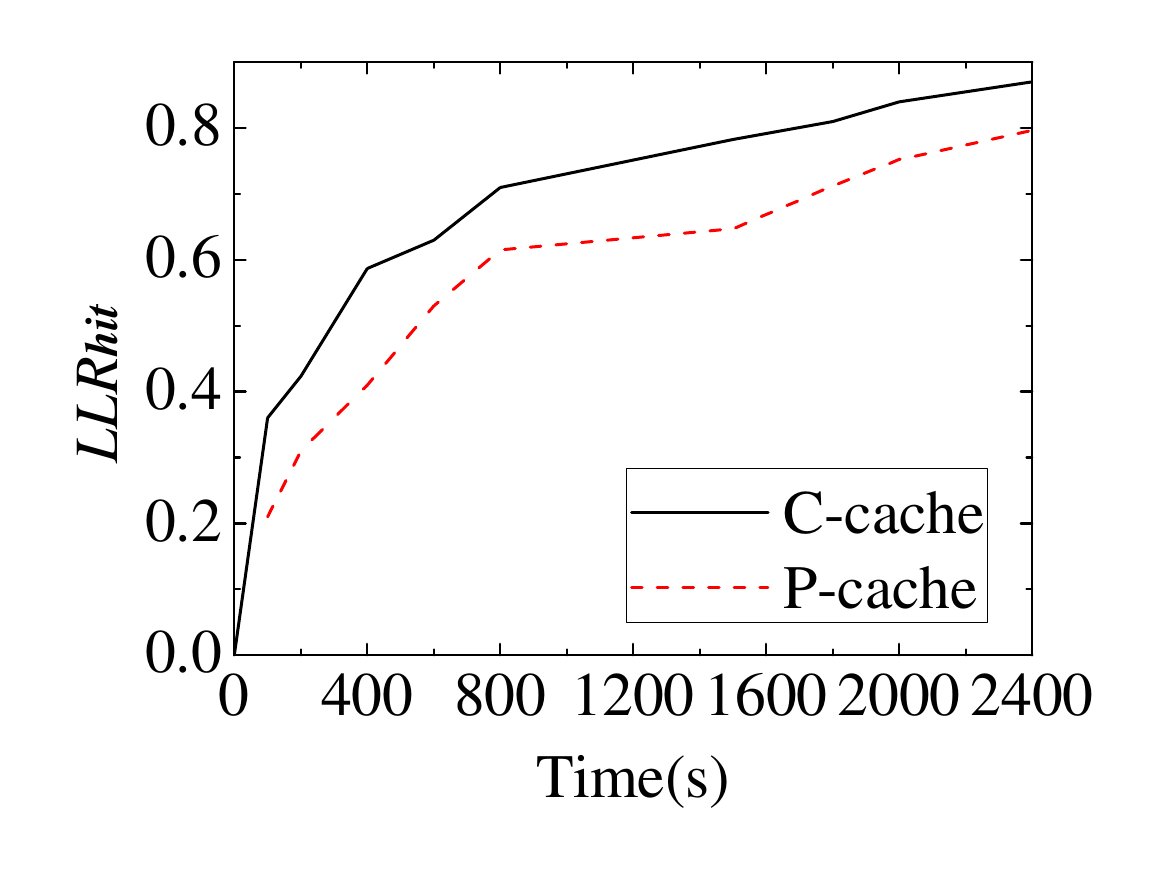}}
	\caption{Local hit ratio of VGG}
	\label{Local hit ratio2}
\end{figure}

\begin{figure}
	\centering
	\subfigure[$GLR_{hit}$ during training MLP on D1]{\includegraphics[width=0.36\textwidth]{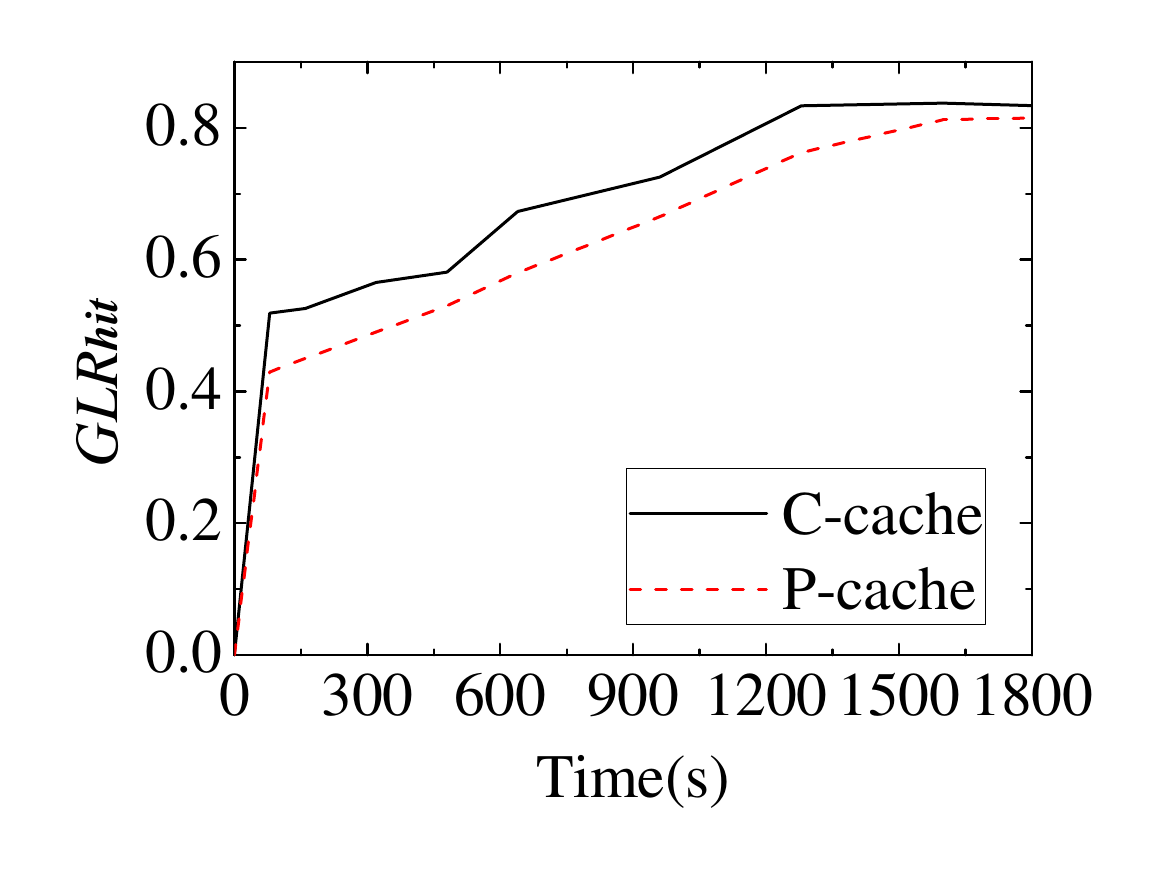}}
	\subfigure[$GLR_{hit}$ during training MLP on D2]{\includegraphics[width=0.36\textwidth]{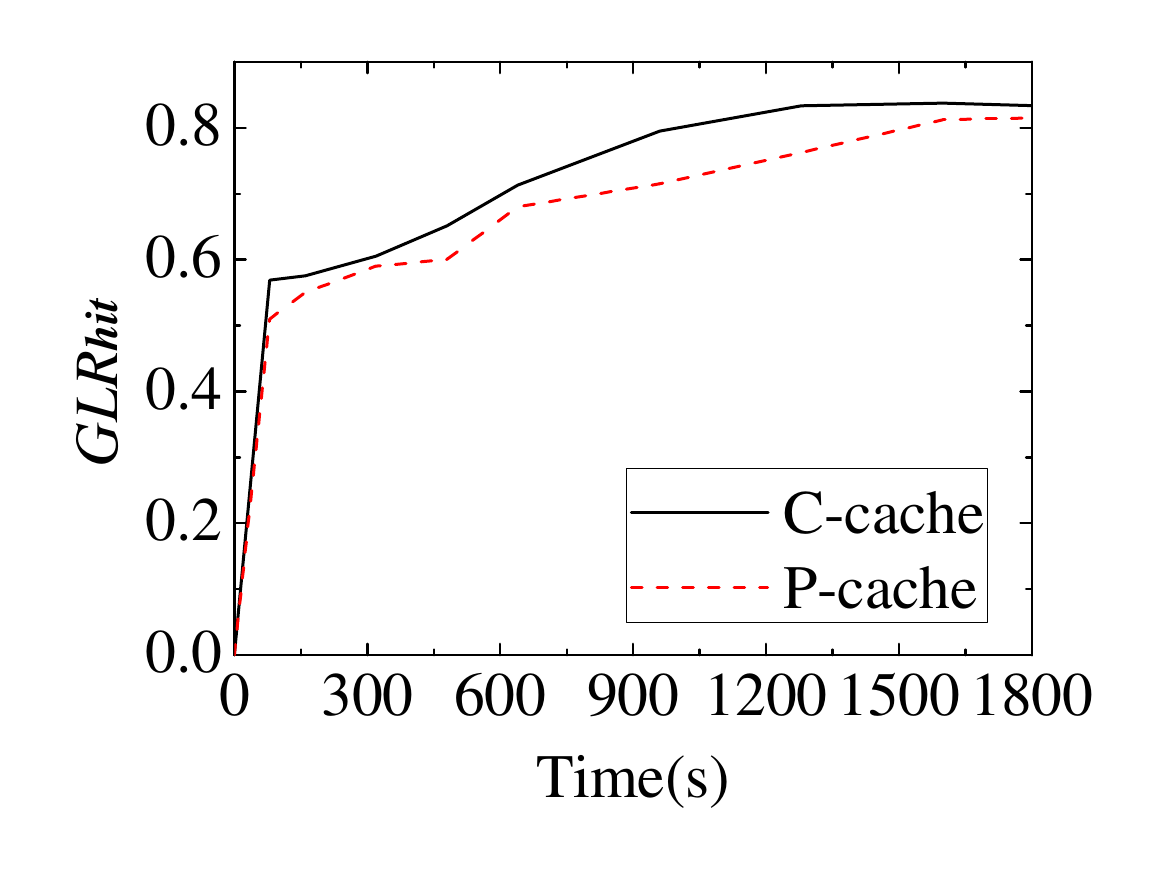}}
	\caption{Global learning hit ratio of MLP}
	\label{Global learning hit ratio1}
\end{figure}

\begin{figure}
	\centering
	\subfigure[$GLR_{hit}$ during training VGG on D3]{\includegraphics[width=0.36\textwidth]{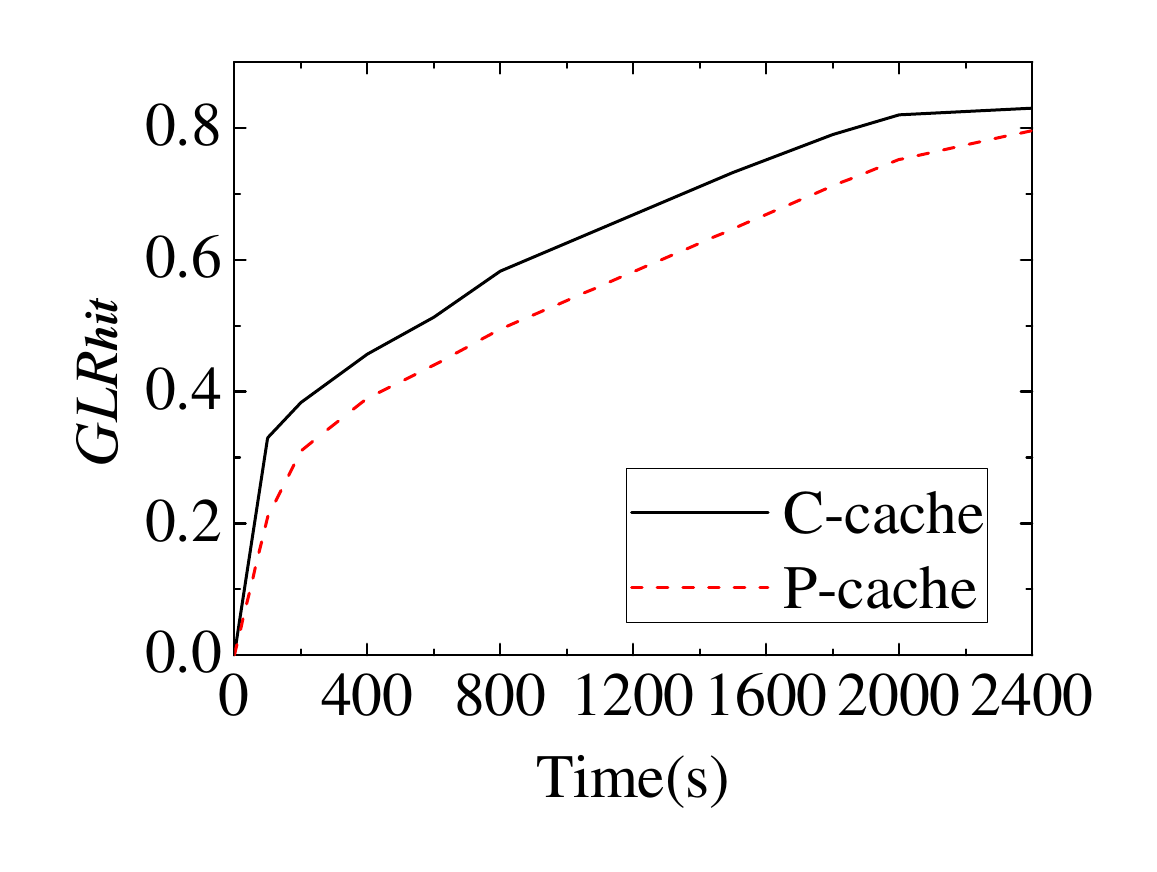}}
	\subfigure[$GLR_{hit}$ during training VGG on D4]{\includegraphics[width=0.36\textwidth]{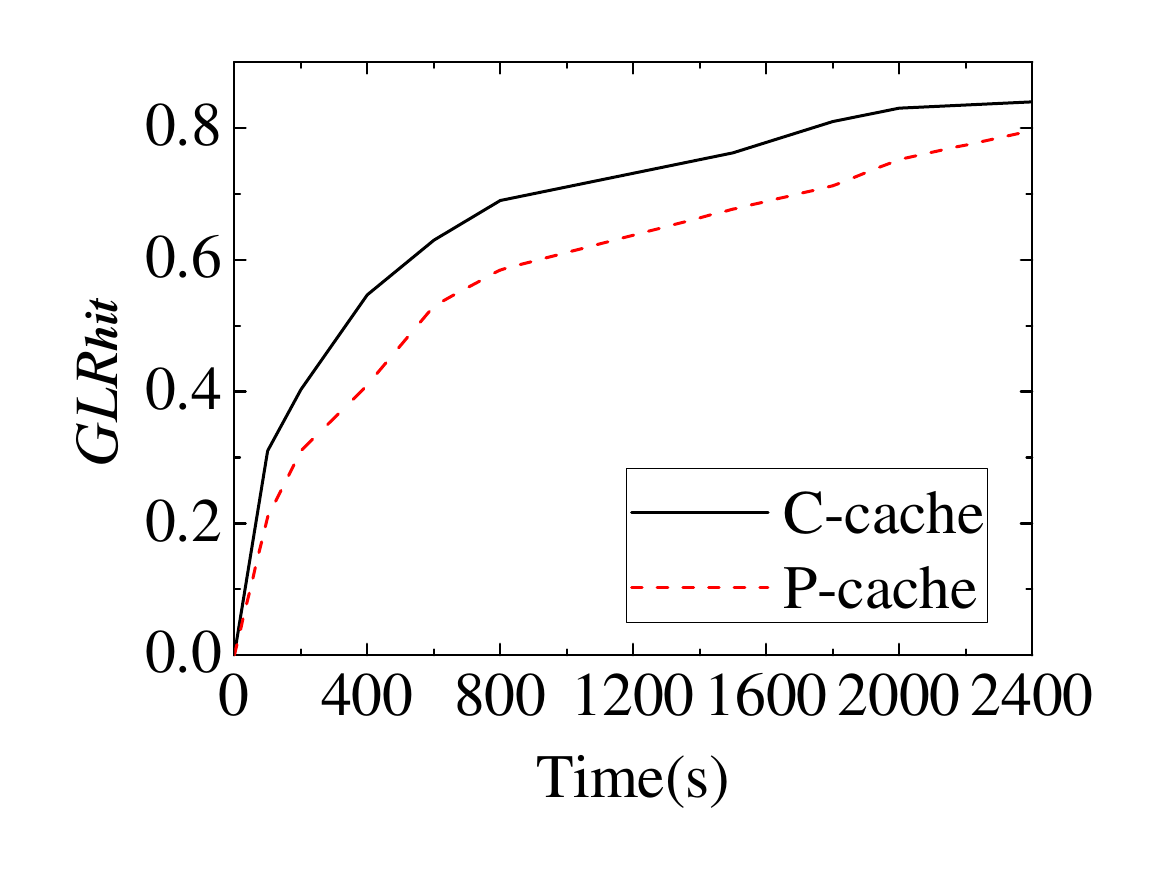}}
	\caption{Global learning hit ratio of VGG}
	\label{Global learning hit ratio2}
\end{figure}

Since the Centralized scheme trains models in the data center without caching data in edge nodes, we only compare the cache hit ratio of the baseline, P-cache, and the proposed C-cache. Local learning hit ratio is depicted in Figure~\ref{Local hit ratio1} and Figure~\ref{Local hit ratio2}. Global learning hit ratio is depicted in Figure~\ref{Global learning hit ratio1} and Figure~\ref{Global learning hit ratio2}. The local learning hit ratios of C-cache and P-cache increase to their maximum stable value of 0.87 and 0.85. The global learning hit ratios of C-cache and P-cache increase to their maximum stable value of 0.83 and 0.81, while the learning data are generated and cached in different edge nodes.

\begin{figure}
	\centering
	\subfigure[$R_{hit}$ during training MLP on D1]{\includegraphics[width=0.36\textwidth]{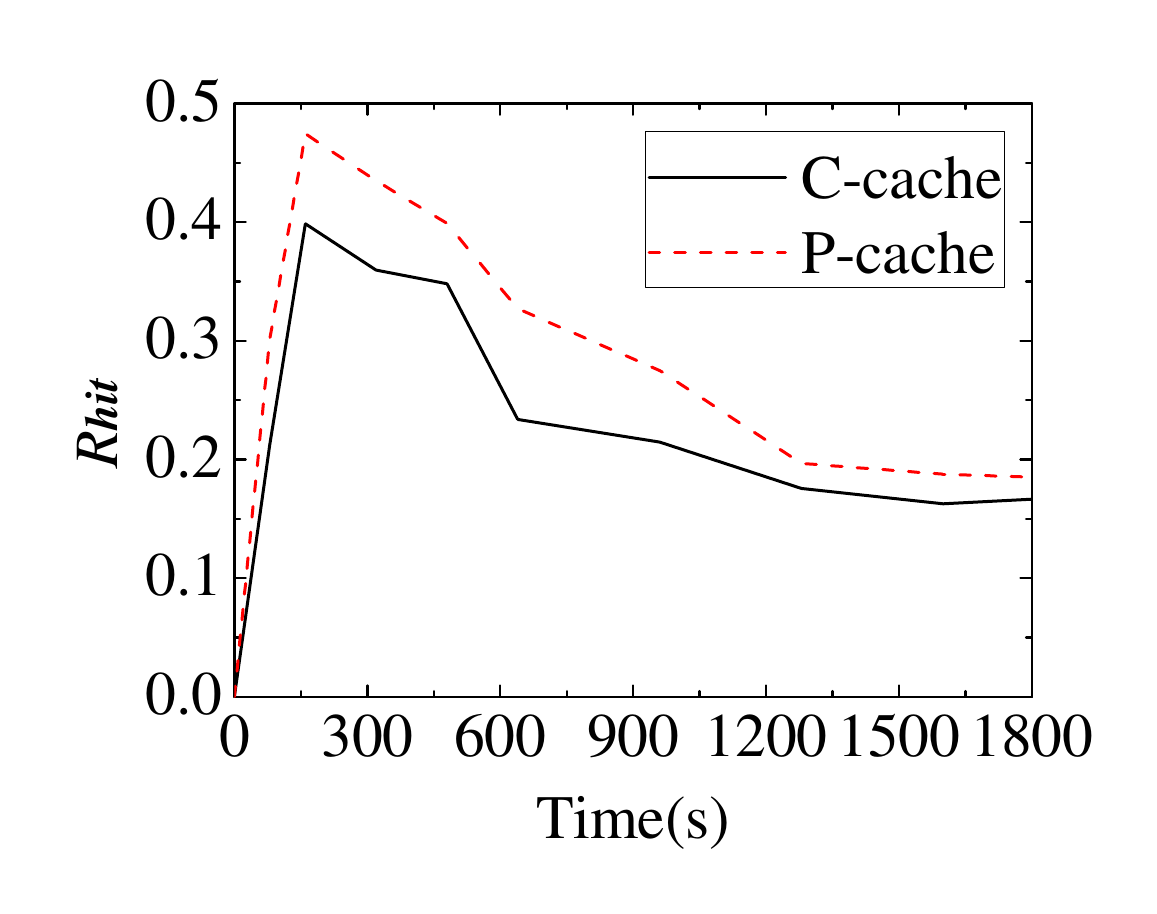}}
	\subfigure[$R_{hit}$ during training MLP on D2]{\includegraphics[width=0.36\textwidth]{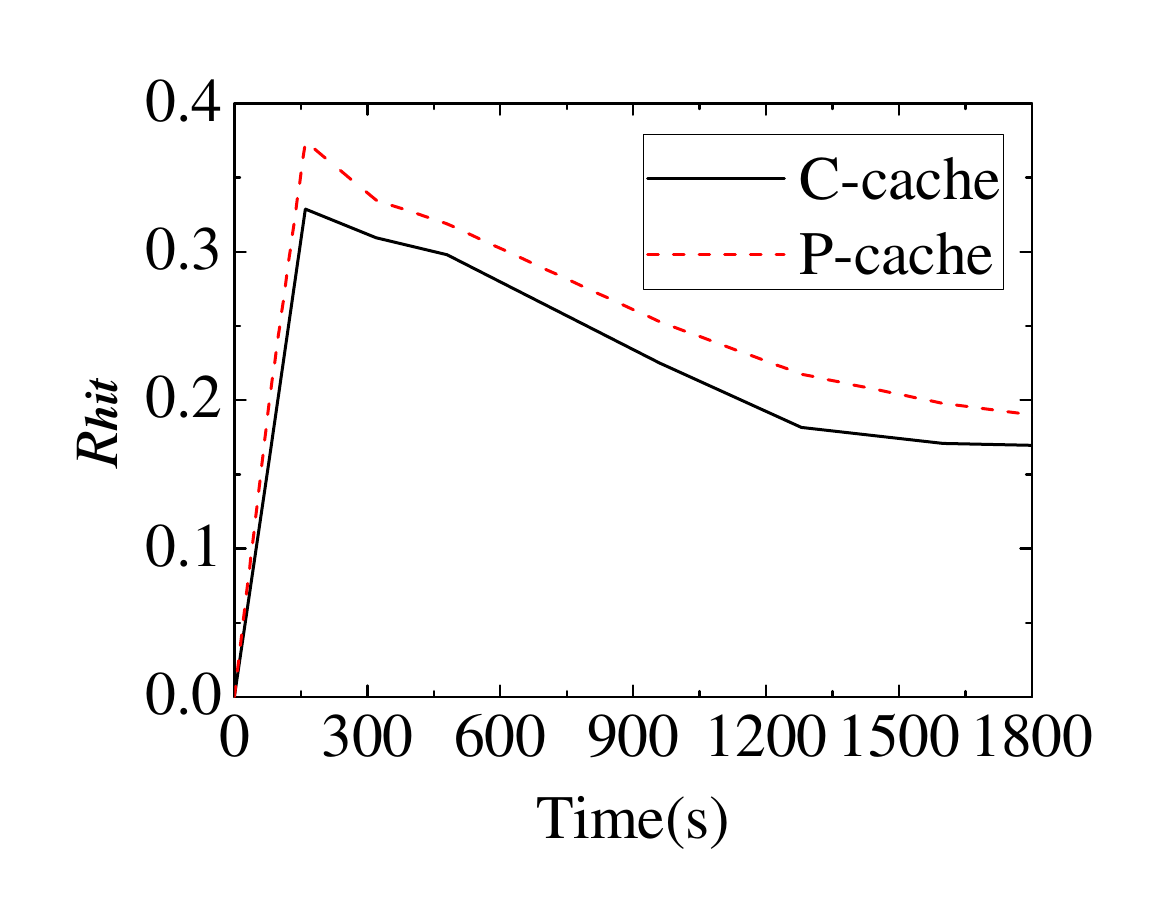}}
	\caption{Global background hit ratio of MLP}
	\label{Global background hit ratio1}
\end{figure}

\begin{figure}
	\centering
	\subfigure[$R_{hit}$ during training VGG on D3]{\includegraphics[width=0.36\textwidth]{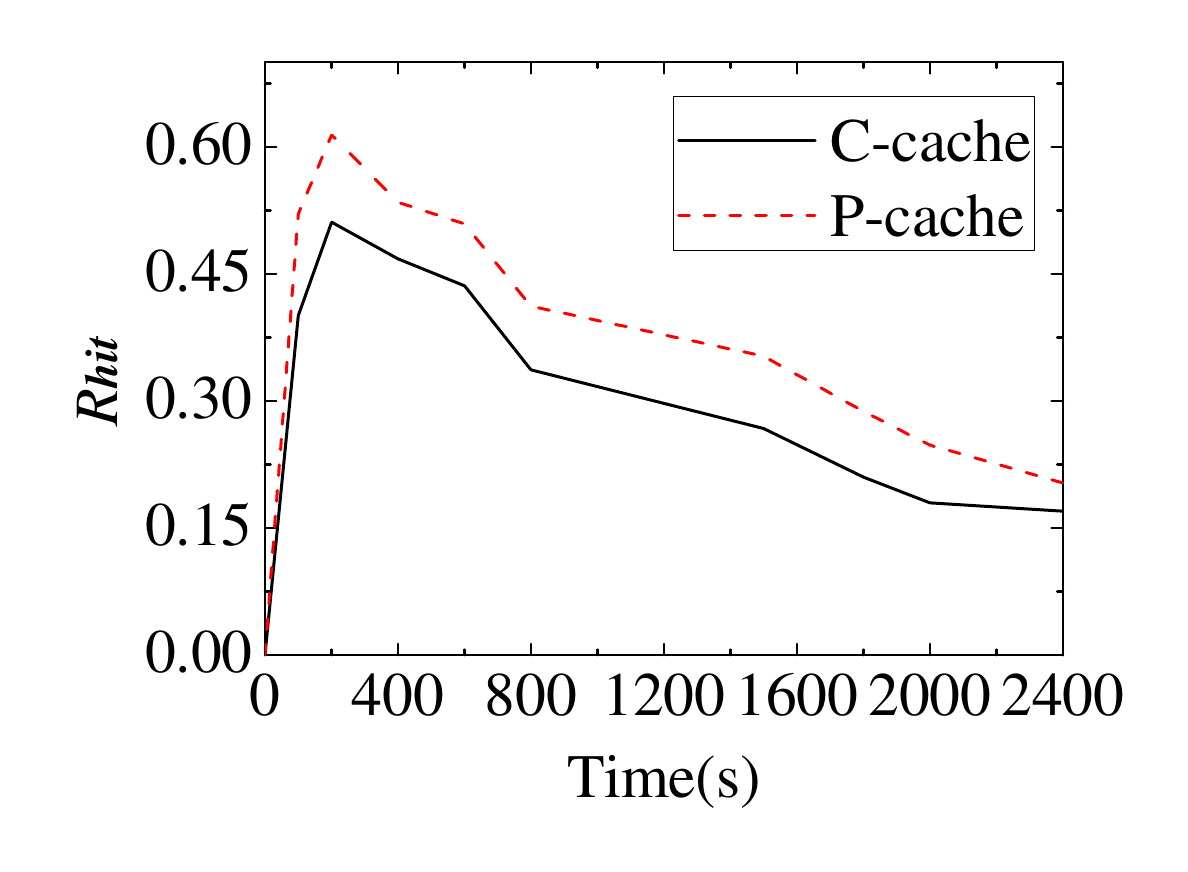}}
	\subfigure[$R_{hit}$ during training VGG on D4]{\includegraphics[width=0.36\textwidth]{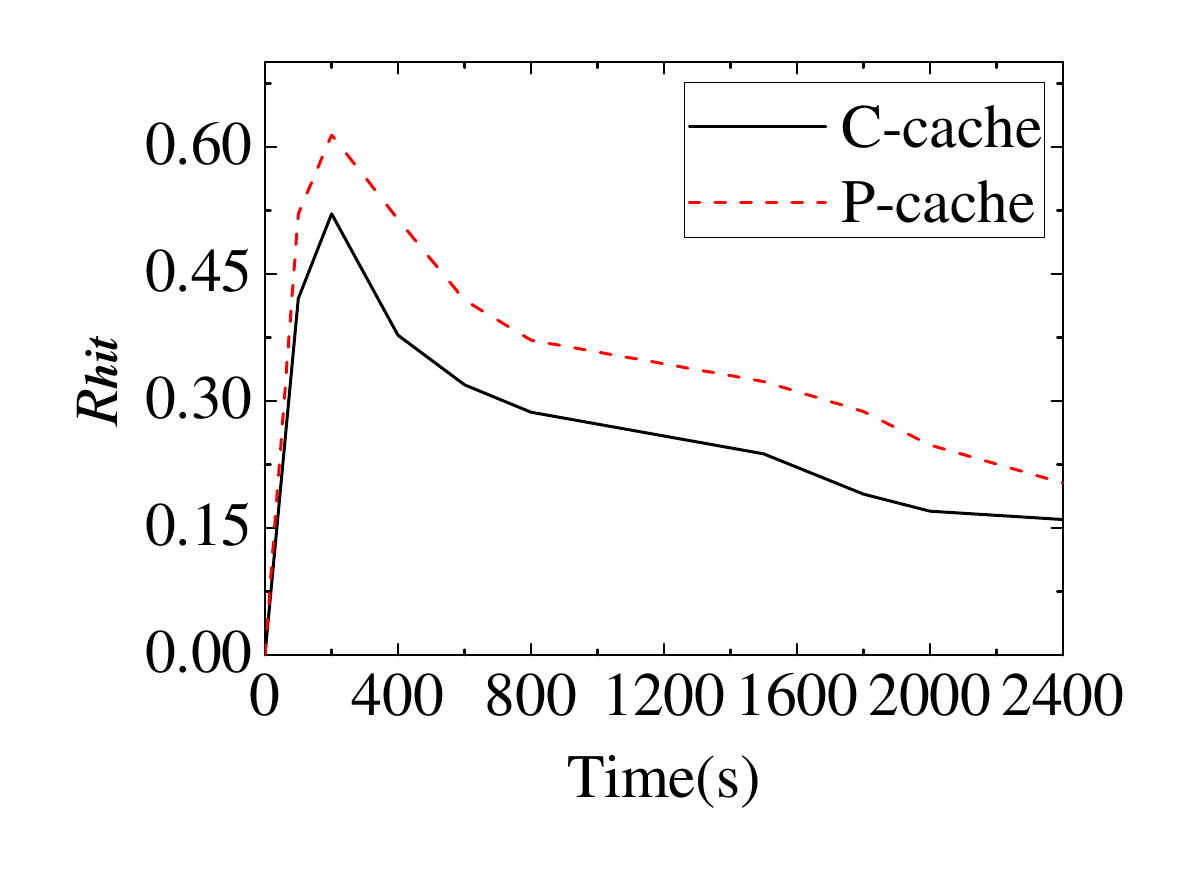}}
	\caption{Global background hit ratio of VGG}
	\label{Global background hit ratio2}
\end{figure}

Figure~\ref{Global background hit ratio1} and Figure~\ref{Global background hit ratio2} are depicted hit ratio of background traffic data. The cache hit ratio of background traffic data first increases over time, and when the learning data increases, more background traffic data are switched out from the caches of edge computing nodes. Consequently, the cache hit ratios of background traffic data in C-cache and P-cache decrease to 0.17 and 0.19. Regarding different training models and datasets, the cache hit ratio under C-cache declines faster than that under P-cache. This is because C-cache can use learning data better than P-cache, and less available cache space is reserved for caching background traffic data.

\subsubsection{Transmission overhead and Learning Latency}
\begin{figure}[htbp]
	\centering
	\includegraphics[width=0.48\textwidth]{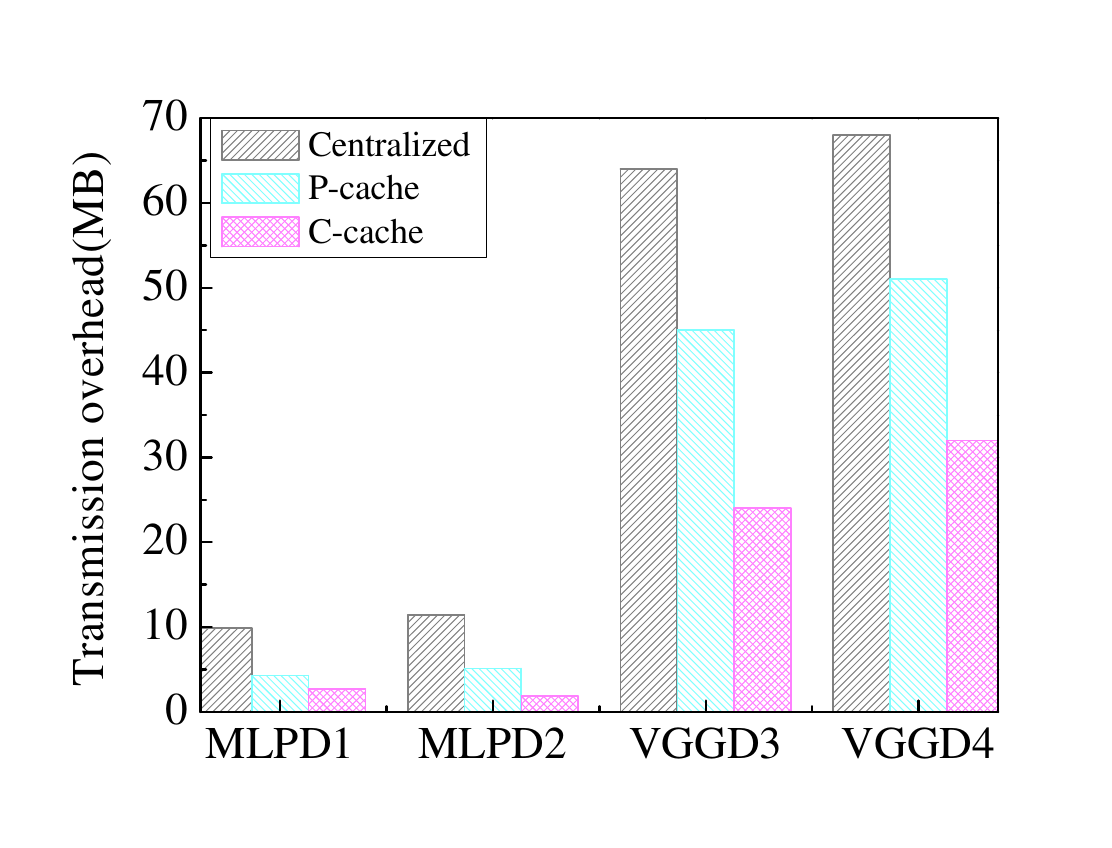}
	\caption{Transmission overhead}
	\label{Transmission overhead}
\end{figure}

To evaluate the communication and time overhead of the proposed scheme, we compare two baselines, Centralized and P-cache with our C-cache in terms of transmission overhead and learning latency. Among them, the transmission overhead is depicted in Figure~\ref{Transmission overhead}. No matter which models or data sets are taken into consideration, C-cache always has the least transmission overhead. More powerful model like VGG will consume more communication resource, while the transmission overhead of the Centralized scheme is twice as much as that of C-cache. All learning data needs to be sent to the data center, which makes the Centralized scheme has the largest transmission overhead. In addition, the rational data requests and collaborative caching benefits C-cache regarding the transmission overhead. Valuable data are cached on edge nodes, and thus reduce redundant data transmission among different edge nodes.

\begin{figure}[htbp]
	\centering
	\includegraphics[width=0.48\textwidth]{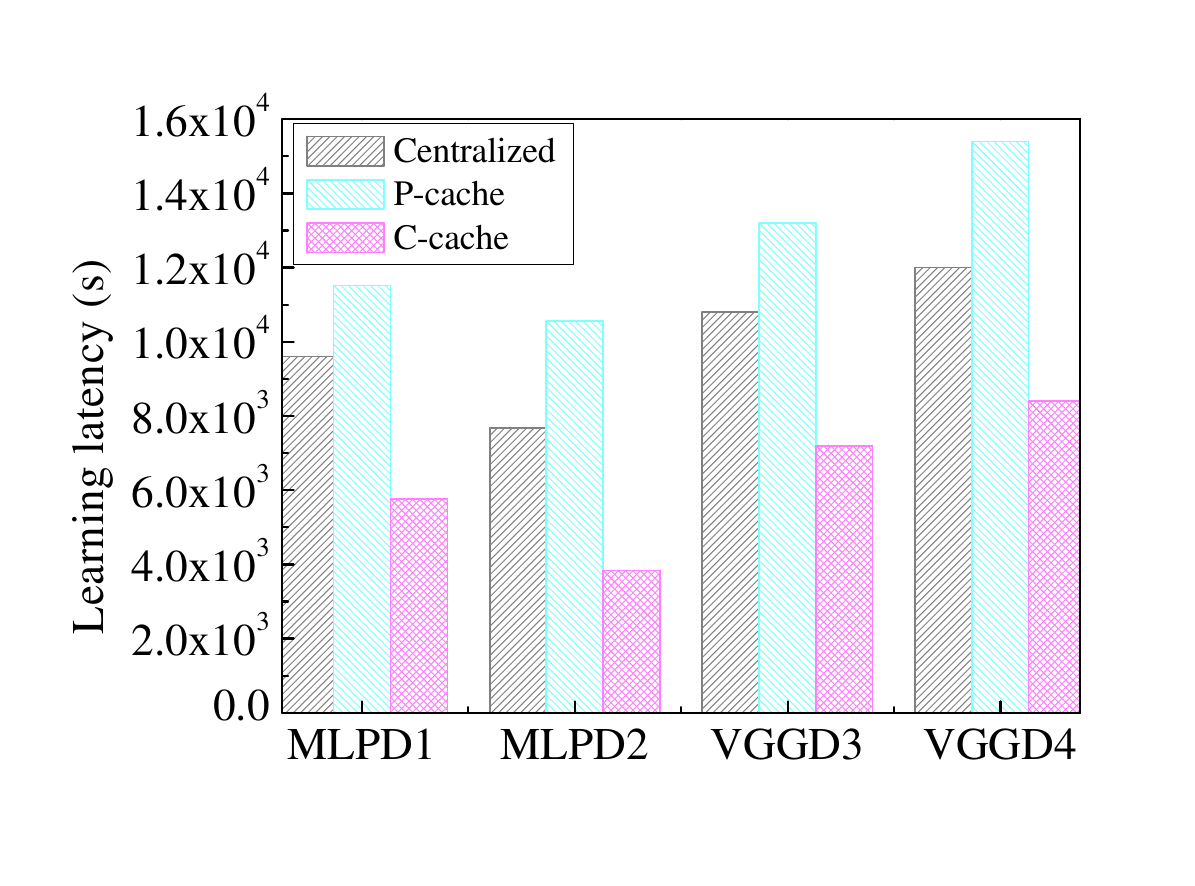}
	\caption{Learning latency}
	\label{Learning latency}
\end{figure}

Figure~\ref{Learning latency} depicts the learning latency of different models on the three schemes. Both MLP and VGG can take advantage of the collaborative caching of C-cache to achieve fast convergence. The maximum difference between learning latency on P-cache and C-cache is 7000s.  Within one or two hours, C-cache provides sufficient data items for sub-model learning and their ensemble process. 
Since the Centralized scheme collect all of the training data to support model training, the model learning latency on Centralized is less than P-cache. On the other hand, large transmission latency of Centralized also degrades the model learning efficiency on Centralized.

\subsubsection{Learning Accuracy}
\begin{table}[]
	\centering
	\begin{tabular}{|l|c|c|c|c|}
		\hline
		\multirow{2}*{Scheme} & \multicolumn{2}{c|}{MLP}&\multicolumn{2}{c|}{VGG} \\
		\cline{2-5}
		& D1 & D2 & D3 & D4\\
		\hline
		Centralized& 
		\makecell{\textbf{0.848}} & 
		\makecell{\textbf{0.968}} & 
		\makecell{\textbf{0.917}} & 
		\makecell{\textbf{0.923}} \\
		P-cache& 0.789 & 0.947 & 0.827 & 0.852\\
		\makecell{C-cache}& 
		\makecell{0.847} & 
		\makecell{\textbf{0.968}}&
		\makecell{\textbf{0.917}} &
		\makecell{\textbf{0.923}}  \\
		\hline
	\end{tabular}
	\caption{Learning accuracy comparison}
	\label{accuracy}
\end{table}

Table~\ref{accuracy} depicts the accuracy of MLP and VGG models trained on different schemes. Regarding different training models and datasets, the C-cache and the Centralized scheme achieve the similar high performance in accuracy. This is because C-cache can provide more valuable training data to support model training, while the Centralized scheme collect all the training data to support model training. On the contrary, P-cache cannot provide sufficient training data in a short time period, which effect the sub-models training on edge nodes and thus degrades the performance of the ensemble results.

\section{CONCLUSION}
\label{sec:con}
In this paper, we propose an adaptive in-network collaborative caching scheme to support efficient ensemble learning at edge. 
In this scheme, edge nodes collaborate to obtain cached data items with different features as much as possible, train sub-models with large differences, thus effectively improving the performance of ensemble learning. 
The extensive simulations demonstrate that our proposed collaborative caching scheme in edge network can significantly reduce learning latency and transmission overhead for ensemble learning.

\bibliographystyle{unsrt}  
\bibliography{main}  

\begin{thebibliography}{10}

\bibitem{deng2020edge}
Shuiguang Deng, Hailiang Zhao, Weijia Fang, Jianwei Yin, Schahram Dustdar, and
  Albert~Y Zomaya.
\newblock Edge intelligence: the confluence of edge computing and artificial
  intelligence.
\newblock {\em IEEE Internet of Things Journal}, 2020.

\bibitem{zhou2019edge}
Zhi Zhou, Xu~Chen, En~Li, Liekang Zeng, Ke~Luo, and Junshan Zhang.
\newblock Edge intelligence: Paving the last mile of artificial intelligence
  with edge computing.
\newblock {\em Proceedings of the IEEE}, 107(8):1738--1762, 2019.

\bibitem{chen2019artificial}
Zhuang Chen, Qian He, Lei Liu, Dapeng Lan, Hwei-Ming Chung, and Zhifei Mao.
\newblock An artificial intelligence perspective on mobile edge computing.
\newblock In {\em 2019 IEEE International Conference on Smart Internet of
  Things (SmartIoT)}, pages 100--106. IEEE, 2019.

\bibitem{zhou2009ensemble}
Zhi-Hua Zhou.
\newblock Ensemble learning.
\newblock {\em Encyclopedia of biometrics}, 1:270--273, 2009.

\bibitem{tumer1995theoretical}
Kagan Tumer and Joydeep Ghosh.
\newblock Theoretical foundations of linear and order statistics combiners for
  neural pattern classifiers.
\newblock {\em IEEE Trans. Neural Networks}, 1995.

\bibitem{yuan2018toward}
Quan Yuan, Haibo Zhou, Jinglin Li, Zhihan Liu, Fangchun Yang, and
  Xuemin~Sherman Shen.
\newblock Toward efficient content delivery for automated driving services: An
  edge computing solution.
\newblock {\em IEEE Network}, 32(1):80--86, 2018.

\bibitem{wang2018edge}
Shiqiang Wang, Tiffany Tuor, Theodoros Salonidis, Kin~K Leung, Christian
  Makaya, Ting He, and Kevin Chan.
\newblock When edge meets learning: Adaptive control for resource-constrained
  distributed machine learning.
\newblock In {\em IEEE INFOCOM 2018-IEEE Conference on Computer
  Communications}, pages 63--71. IEEE, 2018.

\bibitem{yang2019federated}
Qiang Yang, Yang Liu, Tianjian Chen, and Yongxin Tong.
\newblock Federated machine learning: Concept and applications.
\newblock {\em ACM Transactions on Intelligent Systems and Technology (TIST)},
  10(2):1--19, 2019.

\bibitem{recht2011hogwild}
Benjamin Recht, Christopher Re, Stephen Wright, and Feng Niu.
\newblock Hogwild: A lock-free approach to parallelizing stochastic gradient
  descent.
\newblock In {\em Advances in neural information processing systems}, pages
  693--701, 2011.

\bibitem{zhang2015deep}
Sixin Zhang, Anna~E Choromanska, and Yann LeCun.
\newblock Deep learning with elastic averaging sgd.
\newblock In {\em Advances in neural information processing systems}, pages
  685--693, 2015.

\bibitem{wen2017terngrad}
Wei Wen, Cong Xu, Feng Yan, Chunpeng Wu, Yandan Wang, Yiran Chen, and Hai Li.
\newblock Terngrad: Ternary gradients to reduce communication in distributed
  deep learning.
\newblock In {\em Advances in neural information processing systems}, pages
  1509--1519, 2017.

\bibitem{yin2017recognition}
Zhong Yin, Mengyuan Zhao, Yongxiong Wang, Jingdong Yang, and Jianhua Zhang.
\newblock Recognition of emotions using multimodal physiological signals and an
  ensemble deep learning model.
\newblock {\em Computer methods and programs in biomedicine}, 140:93--110,
  2017.

\bibitem{kumar2016ensemble}
Ashnil Kumar, Jinman Kim, David Lyndon, Michael Fulham, and Dagan Feng.
\newblock An ensemble of fine-tuned convolutional neural networks for medical
  image classification.
\newblock {\em IEEE journal of biomedical and health informatics},
  21(1):31--40, 2016.

\bibitem{xiao2019svm}
Jianli Xiao.
\newblock Svm and knn ensemble learning for traffic incident detection.
\newblock {\em Physica A: Statistical Mechanics and its Applications},
  517:29--35, 2019.

\bibitem{galicia2019multi}
Antonio Galicia, R~Talavera-Llames, A~Troncoso, Irena Koprinska, and Francisco
  Martnez-lvarez.
\newblock Multi-step forecasting for big data time series based on ensemble
  learning.
\newblock {\em Knowledge-Based Systems}, 163:830--841, 2019.

\bibitem{liu2017ensemble1}
Wei Liu, Miaohui Zhang, Zhiming Luo, and Yuanzheng Cai.
\newblock An ensemble deep learning method for vehicle type classification on
  visual traffic surveillance sensors.
\newblock {\em IEEE Access}, 5:24417--24425, 2017.

\bibitem{liu2017ensemble}
Xiaobo Liu, Zhentao Liu, Guangjun Wang, Zhihua Cai, and Harry Zhang.
\newblock Ensemble transfer learning algorithm.
\newblock {\em IEEE Access}, 6:2389--2396, 2017.

\bibitem{chen2018ensemble}
Xi-liang Chen, Lei Cao, Chen-xi Li, Zhi-xiong Xu, and Jun Lai.
\newblock Ensemble network architecture for deep reinforcement learning.
\newblock {\em Mathematical Problems in Engineering}, 2018, 2018.

\bibitem{amer2020caching}
Ramy Amer, M~Majid Butt, and Nicola Marchetti.
\newblock Caching at the edge in low latency wireless networks.
\newblock {\em Wireless Automation as an Enabler for the Next Industrial
  Revolution}, pages 209--240, 2020.

\bibitem{li2019collaborative}
Chunlin Li, Jianhang Tang, Hengliang Tang, and Youlong Luo.
\newblock Collaborative cache allocation and task scheduling for data-intensive
  applications in edge computing environment.
\newblock {\em Future Generation Computer Systems}, 95:249--264, 2019.

\bibitem{chien2020q}
Wei-Che Chien, Hung-Yen Weng, and Chin-Feng Lai.
\newblock Q-learning based collaborative cache allocation in mobile edge
  computing.
\newblock {\em Future Generation Computer Systems}, 102:603--610, 2020.

\bibitem{ndikumana2017collaborative}
Anselme Ndikumana, Saeed Ullah, Tuan LeAnh, Nguyen~H Tran, and Choong~Seon
  Hong.
\newblock Collaborative cache allocation and computation offloading in mobile
  edge computing.
\newblock In {\em 2017 19th Asia-Pacific Network Operations and Management
  Symposium (APNOMS)}, pages 366--369. IEEE, 2017.

\bibitem{tang2019using}
Jine Tang, Zhangbing Zhou, Xiao Xue, and Gongwen Wang.
\newblock Using collaborative edge-cloud cache for search in internet of
  things.
\newblock {\em IEEE Internet of Things Journal}, 7(2):922--936, 2019.

\bibitem{khan2020lochip}
Junaid~Ahmed Khan, Cedric Westphal, JJ~Garcia-Luna-Aceves, and Yacine
  Ghamri-Doudane.
\newblock Lochip: A distributed collaborative cache management scheme at the
  network edge.
\newblock In {\em NOMS 2020-2020 IEEE/IFIP Network Operations and Management
  Symposium}, pages 1--7. IEEE, 2020.

\bibitem{wei2019automating}
Jinliang Wei, Garth~A Gibson, Phillip~B Gibbons, and Eric~P Xing.
\newblock Automating dependence-aware parallelization of machine learning
  training on distributed shared memory.
\newblock In {\em Proceedings of the Fourteenth EuroSys Conference 2019}, pages
  1--17, 2019.

\bibitem{perrone1993networks}
MP~Perrone and LN~Cooper.
\newblock ``when networks disagree: Ensemble methods for neural networks''in:
  Rj mammone, ed.
\newblock {\em Neural Networks for Speech and Image Processing Chapman Hall},
  1993.

\bibitem{carneiro2010ns}
Gustavo Carneiro.
\newblock Ns-3: Network simulator 3.
\newblock In {\em UTM Lab Meeting April}, volume~20, pages 4--5, 2010.

\bibitem{lopez2014open}
R~Lopez.
\newblock Open nn: An open source neural networks c++ library.
\newblock {\em Artificial Intelligence Techniques, Ltd.: Salamanca, Spain},
  2014.

\bibitem{gama2003accurate}
Jo{\~a}o Gama, Ricardo Rocha, and Pedro Medas.
\newblock Accurate decision trees for mining high-speed data streams.
\newblock In {\em Proceedings of the ninth ACM SIGKDD international conference
  on Knowledge discovery and data mining}, pages 523--528, 2003.

\bibitem{wickramasinghe2016sequence}
Asanga Wickramasinghe, Damith~C Ranasinghe, Christophe Fumeaux, Keith~D Hill,
  and Renuka Visvanathan.
\newblock Sequence learning with passive rfid sensors for real-time bed-egress
  recognition in older people.
\newblock {\em IEEE journal of biomedical and health informatics},
  21(4):917--929, 2016.

\bibitem{schneider2020similarity}
Stefan Schneider, Graham~W Taylor, and Stefan~C Kremer.
\newblock Similarity learning networks for animal individual
  re-identification-beyond the capabilities of a human observer.
\newblock In {\em Proceedings of the IEEE Winter Conference on Applications of
  Computer Vision Workshops}, pages 44--52, 2020.

\bibitem{sarhan2017multimodal}
Shahenda Sarhan, Shaaban Alhassan, and Samir Elmougy.
\newblock Multimodal biometric systems: a comparative study.
\newblock {\em Arabian Journal for Science and Engineering}, 42(2):443--457,
  2017.

\bibitem{zhang2020online}
Mingchuan Zhang, Yangfan Zhou, Wei Quan, Junlong Zhu, Ruijuan Zheng, and
  Qingtao Wu.
\newblock Online learning for iot optimization: A frank-wolfe adam based
  algorithm.
\newblock {\em IEEE Internet of Things Journal}, 2020.

\end{thebibliography}


\end{document}